\documentclass[conference]{IEEEtran} 
\usepackage{graphicx}
\usepackage{enumerate}
\usepackage{amsmath}
\usepackage{amssymb}
\usepackage{natbib}
\usepackage[utf8]{inputenc}
\usepackage{fancyhdr}
\usepackage{epstopdf}
\usepackage{color}

\newcommand{\W}{W}
\newcommand{\bfW}{\mathbf{W}}
\newcommand{\Wcal}{\mathcal{W}}
\newcommand{\Bcal}{\mathcal{B}}
\newcommand{\Ucal}{\mathcal{U}}

\newcommand{\url}{U}
\newcommand{\hw}{\mathbf{\hat{W}}}
\newcommand{\wh}{\mathbf{\hat{W}}}
\newcommand{\dk}{\mathbf{\mathfrak{d}}}

\newcommand{\bfD}{\mathbf{\mathfrak{D}}}
\newcommand{\bfa}{\mathbf{a}}
\newcommand{\bfalpha}{\mathbf{\alpha}}
\newcommand{\bfdelta}{\mathbf{\delta}}
\newcommand{\bfC}{\mathbf{C}}

\newcommand{\bfr}{\mathbf{r}}

\newcommand{\B}{\mathcal{B}}

\newcommand{\ajout}[1]{\textcolor{red}{[]}}

\begin{document} 

\pagestyle{fancy}
\fancyfoot[L]{\small{978-0-7695-3823-5/09 \$26.00 © 2009 IEEE \\
DOI 10.1109/CSE.2009.105 114}}
\setcounter{page}{114}
\fancyhead[C]{2009 International Conference on Computational Science and Engineering }
\renewcommand{\headrulewidth}{0pt}

\title{Socio-semantic dynamics in a blog network}
\author{\IEEEauthorblockN{\bf Jean-Philippe Cointet}
\IEEEauthorblockA{CREA  \& TSV\\
CNRS-Ecole Polytechnique \& INRA\\
ISC - 57-59, rue Lhomond\\
F-75005 Paris, France\\
\texttt{cointet@shs.polytechnique.fr}}
\and
\IEEEauthorblockN{\bf Camille Roth}
\IEEEauthorblockA{CAMS \\
CNRS-EHESS\\
54, bd Raspail\\
F-75006 Paris, France\\
\texttt{roth@ehess.fr}}
}

\maketitle 
\thispagestyle{fancy}

\begin{abstract}
The blogosphere can be construed as a knowledge network made of bloggers who are interacting through a social network to share, exchange or produce information. We claim that the social and semantic dimensions are essentially co-determined and propose to investigate the co-evolutionary dynamics of the blogosphere by examining two intertwined issues: first, how does knowledge distribution drive new interactions and thus influence the social network topology? Second, which role structural network properties play in the information circulation in the system?
We adopt an empirical standpoint by analyzing the semantic and social activity of a portion of the US political blogosphere, monitored on a period of four months.
\end{abstract}




\section{The ``blogosphere'' as a\\socio-semantic system}\smallskip

The blogosphere essentially gathers individuals who share, exchange and produce information and interact online by posting comments or referencing each other. As such, it is a socio-semantic network, in the sense that each blog can be characterized both by a relational profile, determined by its position in the underlying  social network, and by a semantic profile, which describes cognitive attributes. 

Adopting a dual perspective on these knowledge networks is likely to provide a better knowledge of the key mechanisms underlying their organization and evolution: essentially, both dimensions co-evolve, for instance network dynamics is likely to be affected by the distribution of knowledge, if we assume that semantic homophily is a driving force behind  network evolution. Structural features of the implicit social network may also give rise to some specific patterns regarding knowledge distribution. Put differently, by supporting diffusion processes social networks may diversely affect information circulation among bloggers. 

We propose to investigate empirically the coevolutionary dynamics of a portion of the blogosphere by examining the two intertwined following issues: 
\begin{enumerate}[(i)]
\item how does knowledge distribution influence new relationship appearance, thereby influencing the topology?
\item how, in turn, do structural network properties  play a role in the way information circulates in the system?
\end{enumerate}


\subsection*{Related work}
Blogs attracted much attention as an empirical goldmine for quantitative social science and, more theoretically, as a rich instance of social and semantic complex system \cite{adar:impl,Cohen:2006p282,Glance:2004p2283,herr:conv,1035162,Leskovec:2007p186,Leskovec:2007p1939,shi1001lbt}. This recent effort is part of a broader  interest in online knowledge-based networks, including for instance wikis \cite{vieg:talk} or content-sharing websites such as Flickr \cite{marl:posi}, which fundamentally are virtual spaces dedicated to production, sharing, and circulation of opinion, multimedia resource and more broadly information; and where various kinds of social interactions and collaborations are channeled by so-called ``web 2.0'' technologies.\\
Political blogging itself is also the focus of a decent part of the literature in that it allows investigation of multiple current issues, including influence of bloggers over media coverage or over the general political debate \cite{adam:poli,coin:inte,Drezner:2004p2285,Wallsten:2005p2284}. 

Many of these studies focus on the \textit{blogosphere} or  \textit{blogspace} with a social network perspective, aiming at measuring and characterizing topological properties including link configurations, cohesiveness phenomena and existence of groups or communities \cite{Java:2006p1951,Kumar:2005p75,McGlohon:2007p1888}. Beyond a strictly structural approach, static descriptions of the joint distribution of topics and social configuration of a blogosphere has been achieved by \cite{adam:poli}; however the dynamic interrelations of these two dimensions remains a current problem.\\
Further, studies considering the blogosphere as an informational system have mainly focused on investigating topic and opinion evolution --- thanks to the fine-grained dynamics of the underlying data --- thereby developing automatic trend detection methods \cite{Glance:2004p2283}, characterizing opinion dynamics using sentiment analysis \cite{Mishne:2006p2282}, or exploring the coexistence of chatters and spikes in blog conversations, and their cyclic behaviors \cite{Balog:2006p2268,Bansal:2007p2270,Gruhl:2004p79}, {\em inter alia}.

\smallskip
Blogs and more generally Internet-based communication systems have provided a novel opportunity for diffusion studies, through the in-vivo observation of what is generally refered to as ``cascade dynamics'' \cite{Kempe:2003p180,Kleinberg:2007p1971}.\\
{This feature is common notably to viral marketing studies  on large-scale online datasets exhibiting diffusion phenomena; including 
\cite{Leskovec:2007p186} which explores the distribution of the probability of purchasing a cultural consumer good when a large on-line retailer user receives a certain number of recommendations sent by her friends; and \cite{back:grou} computes the probability for one to join a Livejournal community when she already has some friends in it.
More specifically cascades in blog networks have been extensively described by considering chains of posts citing each other as information pathways \cite{Leskovec:2007p1939,Leskovec:2007p1889,McGlohon:2007p1888}. In these cases as well as in other studies not restrained to blog networks  \cite{Iribarren:2007p1892} the focus has been put on influence spread through the study of the \emph{topological} properties of cascades (such as typical patterns of cascade, distribution of cascade sizes, etc.)} 

\smallskip
Eventually, little is known yet on the dynamic underpinnings of content distribution over agents with respect to topology and on the processes underlying the actual formation of heterogeneous topical communities; or, more broadly, on the very intertwining of social and semantic dimensions and their effect on information propagation, with the notable exception of \cite{adar:impl}.\\
Most often, one only of the social and the semantic dimensions is considered. 
On one hand indeed, link creation patterns are generally essentially appraised through structural attributes rather than cognitive/semantic properties of blogs. As for diffusion, either \emph{content} evolution is studied independently of the topology, or topology is the only reference frame for diffusion (one observes the propagation of links \emph{of} the social network \emph{along} the social network --- \hbox{i.e.} some sort of structural transitivity). 
On the other hand, endeavors at understanding what triggers or increases diffusion have given a prevailing role to ego-centered characterizations (i.e. diffusion is often seen as stemming from individual properties, rather than the shape of the network at large).

\smallskip
{Put shortly, in terms of diffusion, taking into account both the network structure and a transmission process on objects \emph{distinct} of this structure is so far a current challenge. In this respect, we also aim at 
assessing how actual content diffusion pathways can be correlated with the (mostly distinct) underlying social network that supports such information circulation.}

\subsection*{Outline} The paper is organized as follows: in the next section we first introduce the empirical protocol. Section~\ref{sec:topology} focuses on the dynamics of link creation in the comment and post networks according to both structural and semantic features, while Sec.~\ref{sec:diffusion} investigates the dynamics of information propagation according to the underlying topology.

\fancyhead{}
\fancyfoot{}

\section{Experimental framework}\label{sec:protocol}\smallskip

\subsection{A bounded subset of the US blogosphere}
Our study is based on the observation of the activity of a medium-sized yet topically well-bounded portion of the US political blogosphere which has been gathered by {\sc Linkfluence} under the ``PresidentialWatch08'' project.\footnote{http://linkfluence.net, http://presidentialwatch08.com}

The dataset consists of $1,066$ blogs, hereafter denoted by $\B$, monitored over the course of four months, from Nov 1, 2007 to Feb 29, 2008. 
For each blogger we crawled the date and full-text content, including hyperlinks, of each post published during the observation period, totaling $71,376$ posts.

\subsection{A dynamic network}

The couple $(\B{},C)$ is the blog network, where $C$ denotes post citation links as an adjacency matrix of size $|\B|\times|\B|$. This data is additionally \emph{dynamic}, with a temporal granularity of one day: we deal with $C_t$, where $t$ ranges from $1$ to $121$: $C_t(i,j)=1$ if $i$ cites $j$ in a post at time $t$; 0 otherwise. 
In the remainder, $t$ may be omitted in the notations when it is implicit. 

We extracted $229,736$ dated edges in $C$, of which $15,032$ are unique (non-repeated links). 
We eventually define an aggregated weighted network as $\bfC_t = \sum_{t'=1}^t C_{t'}$.

\subsection{An epistemic network}
Aside of this structure, content defines a semantic dimension: posts are traditionally dealing with specific issues, sometimes broadcasting particular documents. Although the existence of a clear-cut distinction between high-level topics and specific cultural items may be debated, we assume that (i) textual contents broadly define the various issues a blogger addresses, whereas (ii) explicit URLs (which refer to hyperlinked documents and which are not citation links) define the various specific digital resources a blogger spreads around.

Subsequently, we distinguish:\begin{itemize}
\item a set of high-level topics $\Wcal$ relevantly linked to political commentary in our context, among the most frequent in the corpus (thus excluding rhetorical terms). $\Wcal{}$ is thus made of {$79$} syntagms ranging from names of politicians to issues which kept the blogosphere busy during the presidential campaign, such as ``\textit{climate change}'', ``\textit{national security}'', ``\emph{super Tuesday}'', ``\emph{tax cuts}'', ``\emph{human rights}'', etc.
\item and a set of URLs, noted $\Ucal$, which are not confusable with a link in the citation network --- these are simply online videos, news media article, etc. $\Ucal{}$ is a selection of {$96,637$} URLs (of length larger than 10 characters). Note that these URLs are taken from the limited content of \emph{posts only}, not webpages, so that $\Ucal$ should exclude banners and platform-related links and ads, \emph{inter alia}; it only covers links explicitly cited by bloggers in their posts.
\end{itemize}
More precisely with respect to $\Wcal{}$, we introduce a temporal matrix $\W_t$ which tracks the contents published by bloggers: $\W_t(i,w)$ equals $1$ if term $w\in\Wcal$ appears in a post published on blog $i$ at time $t$, $0$ otherwise.
 Eventually, the $|\Wcal|$-dimensional vector $\bfW_t(i)$ defined as a the sum of rows $\W_{t'}(i)$ for $t'\leq t$
denotes  the aggregation of all topics addressed by blog $i$ until $t$.\\$\bfW_t(i)$ can be seen as the semantic profile of $i$ at $t$.

In a similar fashion for $\Ucal$, we introduce a temporal matrix $\url{}_t$ such that $\url_t(i,u)$ equals $1$ if blog $i$ explicitly refers to a URL $u\in\Ucal$ in a post published at time $t$. Since this matrix will mostly be used for diffusion purposes, we need not define in the present study an aggregated quantity for URL usage.


\section{Evolution of topology:\\The content-based dynamics of link creation}\label{sec:topology}\smallskip

We first study the link creation dynamics with respect to the configuration of the blogosphere, both on a social and semantic level. In particular, we examine the constraint induced by the current socio-semantic network on future citation patterns. 
The structure of both social and semantic configurations may, at least partially, determine link creations. Quite straightforwardly, remoteness in both spaces is likely to modify the landscape of potential relations and, subsequently, modify the likelihood of interaction. In what follows we focus notably on citation propensity with respect to simple notions of topological as well as semantic distances: how do proximity, increased attention or homophily processes actually impact authority attribution in this portion of the blogosphere?

\subsection{Proximity and distance}

To begin with, we define a series of simple distances which are all based on aggregated data at $t$, denoted by ``bold'' notations ($\bfC$ and $\bfW$); that is, we assume each notion of distance between two blogs to depend on the whole history of posting and linking at $t$. 

\subsubsection{Dissimilarity as a semantic distance}\label{sec:semclu}
To  semantically compare a pair of bloggers $i$ and $j$ at $t$, we adopt a classical cosine-based measure of dissimilarity on their profile vectors $\bfW_t(i)$ and $\bfW_t(j)$. We denote this {\bf semantic distance} by $\delta$: concretely, identical profiles yield a $\bfdelta$ of 0, whereas strictly disjoint/orthogonal profiles are separated by a $\bfdelta$ of 1; intermediate values from 0 to 1 indicate increasing levels of dissimilarity.\footnote{To this end, we first need to carry a normalization procedure to weight term occurrences properly, following the  ``{tf}$\cdot${idf}'' canonical approach used  extensively in
 information retrieval, famously introduced in the vector-space model \cite{salt:vect}. This approach more precisely  consists in weighting the ``term frequency'', ``tf'' (so that  most used terms in a given blog are more important) with the so-called ``inverse document frequency'', ``idf'', or frequency of the term in the whole corpus of blogs (so that rarer terms in the blogosphere are weighted more: this takes into account the discriminating power of terms which, while usually rare in the corpus, are  being abnormally mentioned by a given blog).

For this computation, profiles $\bfW_t(i)$ are thus actually replaced by tf$\cdot$idf-adjusted profiles $\hw_t(i)$ such that: $$\displaystyle\hw_t(i,w):=
\frac{\bfW_t(i,w)}{\sum_{w=1}^{|\Wcal|} \bfW_t(i,w)}\cdot\log\frac{|\B|}{|\{j, \bfW_t(j,w)>0\}|}$$ where the ``log'' part of the formula is the inverse ratio of the number of blogs where term $w$ appears over the total number of blogs.
Then, we obtain the dissimilarity between blogs $i$ and $j$  by dividing the scalar product of their adjusted profiles by the product of their norm:   
\begin{equation}
\bfdelta_t(i,j)=\displaystyle 1- \frac{\wh_t(i)\cdot\wh_t(j)}{\|\wh_t(i)\|\|\wh_t(j)\|}
\end{equation}
}

\subsubsection{Topological distances}

Because network links are oriented, topological distances will be asymmetric measures, contrarily to the semantic distance $\delta$.

\begin{figure}
\includegraphics[width=2.95cm,height=6.5cm]{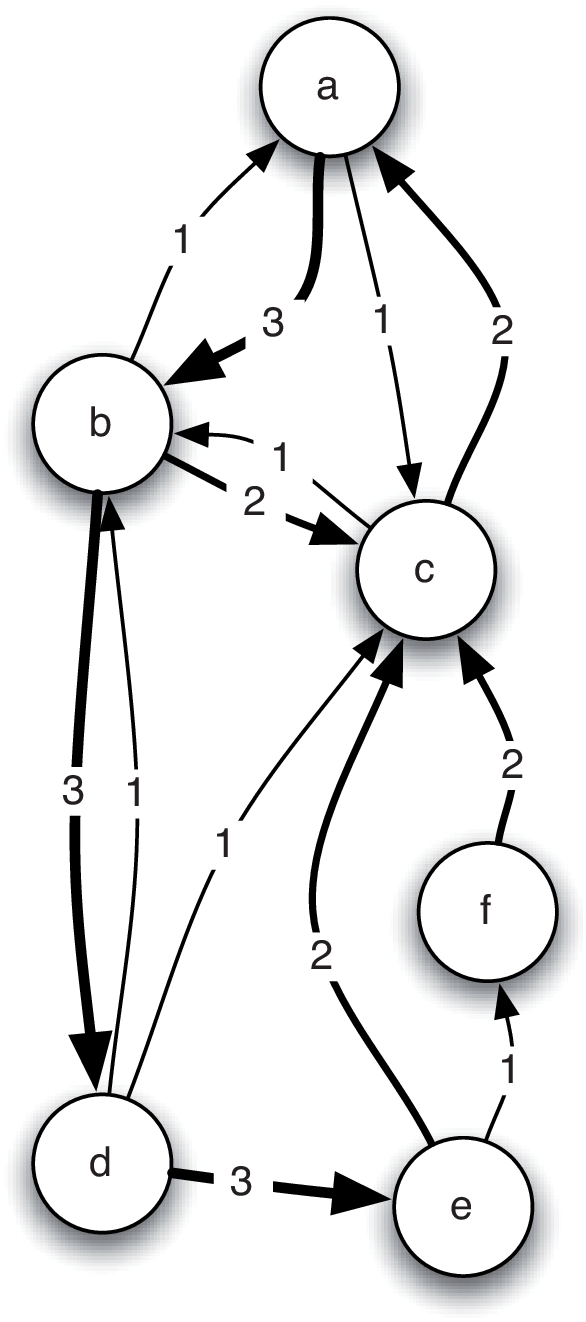}
\vrule
\includegraphics[width=2.95cm,height=6.5cm]{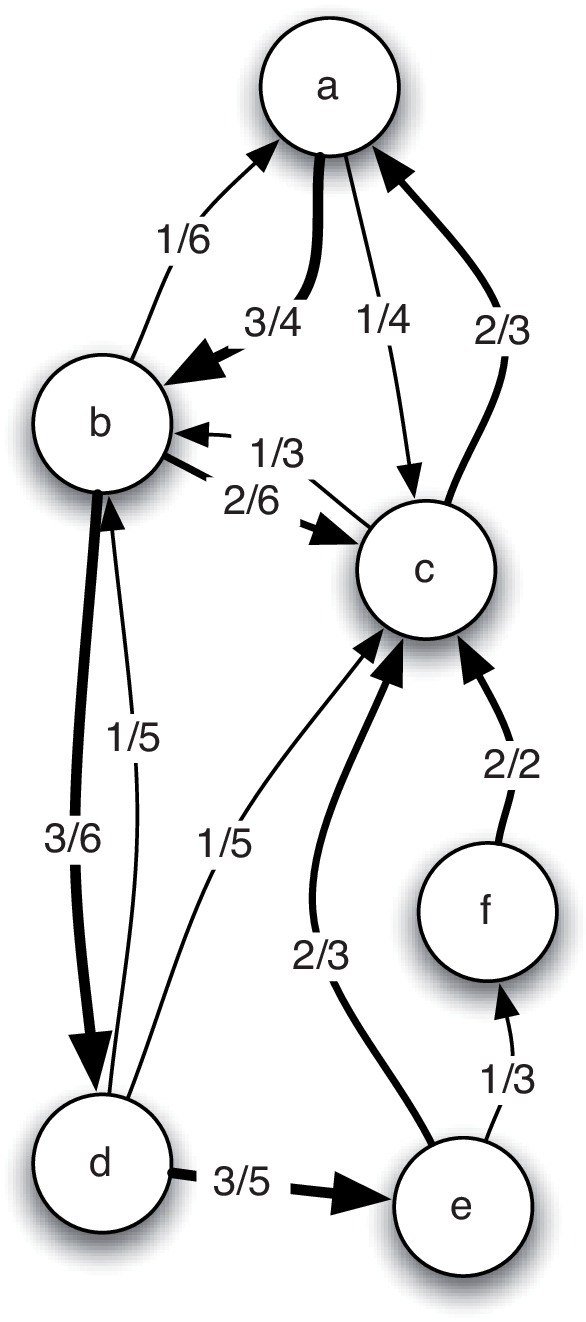}\vrule
\includegraphics[width=2.95cm,height=6.5cm]{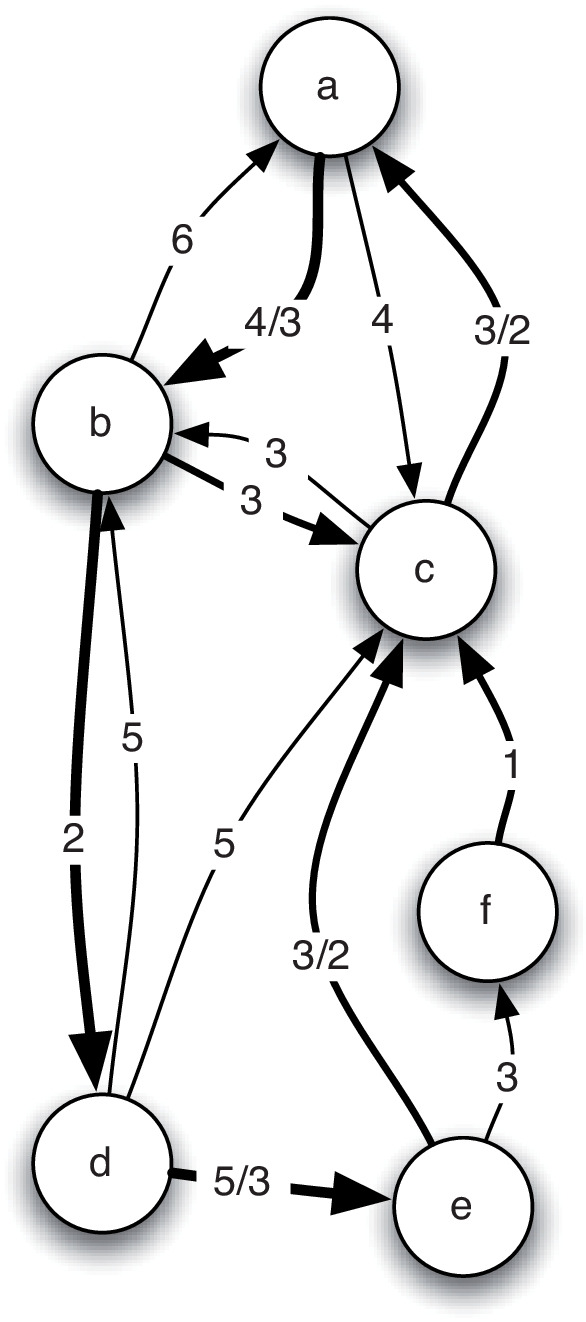}\smallskip\\
$\,{ \, }\quad$\hspace{.4\linewidth}{\scriptsize Attention $\bfa$}\hspace{.18\linewidth}\scriptsize Detachment $\dk$
\caption{{\em Left:} An example of weighted citation network $\bfC_t$: weights trivially correspond to the number of observed links between blogs at some time $t$. {\em Middle} and {\em right:} corresponding attention and detachment values, respectively. \newline
For example, blog $b$ cited $c$ twice out of a total of $1+2+3=6$ citation links, its attention toward $c$ is thus ${\bfa}_t
({b,c})=\frac{2}{6}$. \emph{Detachment} $\dk_t(b,c)$ equals the inverse of the attention from $b$ to $c$, it is $3$. \emph{Detachment-based distance} $\partial_t(b,c)$ is also $3$ (since $b-c$ is the shortest weighted path from $b$ to $c$), while $\partial_t(b,e)=2+5/3=11/3$.} 
\label{fig:attention}
\end{figure}

We first classically define the \textbf{social distance} $d_t(i,j)$ between two blogs $i$ and $j$ in $\bfC$ as the length of the shortest path linking $i$ to $j$ in that network, irrespective of link  weights. This basically refers to the number of steps one has to follow to reach another blog. On the example of Fig.~\ref{fig:attention}-left, $d_t(b,f)=3$.

\medskip
Since influence effects relate to attentional features, we suggest that a notion of remoteness based on ``attention'' may also be relevant. In this respect, we define a dyadic attention $\bfa_t$ by normalizing every row of $\bfC_t$: $${\bfa}_t({i,j})=\frac{\bfC_t(i,j)}{\sum_{j=1}^{|\B|}\bfC_t(i,j)}$$
${\bfa}_t({i,j})$ is thus simply the proportion of links going from $i$ to $j$ among all outgoing links from $i$. Higher values indicate higher focus by $i$ on $j$. Note that a similar notion is called ``influence matrix'' in \cite{Java:2006p1951}.

Now, we can define an opposite notion to attention by considering inverse values of $\bfa$, defining a measure of \emph{detachment} as 
$\dk(i,j)=\displaystyle\frac{1}{\bfa(i,j)}$.
In other words, $\dk(i,j)$ can be compared with a relative cost for information to reach $i$ directly from $j$. It is equal to infinity if there is no link from $i$ to $j$, it is decreasing when attention of $i$ towards $j$ is growing. Basically, for instance, if $i$ has three times more links towards $j$ than towards $k$, then $i$'s detachment to $j$ is three times lower.

Eventually, we define a {\bf detachment-based distance} as the minimal weighted distance \cite{dijkstra1959ntp} in a weighted graph $\bfD$ where link weights from $i$ to $j$ are non-infinite $\dk(i,j)$ values. We denote this detachment-based distance $\partial(i,j)$ --- as such, it can be considered as a measure of attentional remoteness, \hbox{i.e.} lightweight attentional paths will correspond to higher detachment-based distances. See an illustration on Fig.~\ref{fig:attention}.

\subsection{Method for appraising preferential link creation}
While sophisticated regression models have been developed in mathematical social science to measure the preference of link creation \cite{snij:stat}, we stick here to a  basic yet insightful framework for comparing (i) the number of links actually received by some kinds of nodes during a period of time, with (ii) the potential number of such links --- \hbox{i.e.} a kind of ``preferential attachment'' measurement \cite{bara:evol}, here with respect to any kind of property \cite{roth:gene}.  

\newcommand{\dist}{x}
\newcommand{\Dist}{d}
More precisely, 
we define ${f}(\dist)$ as the propensity of formation of new citation links $(i,j)$  such that their social distance is $\Dist(i,j)=\dist$. 
\emph{Put simply, higher propensity values indicate stronger likelihood for dyads at a certain distance  to form, all other things being equal.}

We concretely compute the propensity ${f}(\dist)$ as the proportion of new links appearing in $\bfC$ during a given time period $[t+1,\,t+T]$ and which were at social distance $\dist$ at $t$, among the whole set of possible such pairs at distance $\dist$:
\begin{equation}
{f}(\dist)=\frac{\Big|\Big\{\left.\begin{array}{r}(i,j)\text{ such that }\bfC_{t+T}(i,j)>\bfC_{t}(i,j)\\\text{and }\Dist(i,j)=x\end{array}\right.\Big\}\Big|}{|\{(i,j)\text{ such that }\Dist(i,j)=x\}|}
\end{equation}
Empirically, we estimate various propensities for a series of time steps {$\displaystyle\Big\{[t_k+1,t_k+T]\text{ such that }t_k=60+kT, T=7\Big\}_{k\in\{0,...,7\}}$ --- basically estimating the propensity at a weekly rate, given all previous observed interactions, with the exception that we start the computation only after an initialization period of two months ($t_0=60$).}

Propensities with respect to the social, detachment-based and semantic distances are respectively denoted $f$, $f^\partial$ and $g$. In the figures, all propensities are normalized for comparison purposes.

\begin{figure}[!b]
	\centering
\includegraphics[width=0.85\linewidth,height=5.3cm]{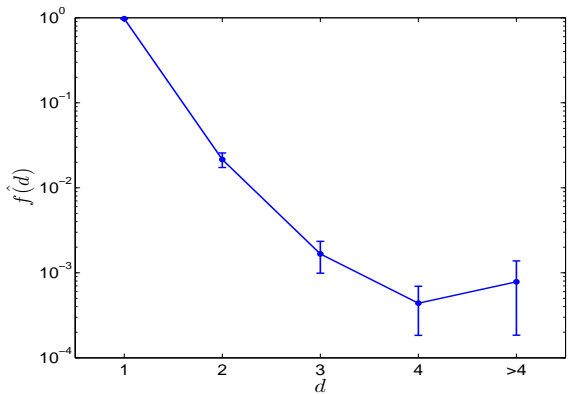}
\caption{Propensity $f$ for new post citation in $\bfC$ as a function of social distance $d$. 
Error bars indicate $95\%$-confidence intervals on means.
\label{prop-soc}}
\end{figure}
\subsection{Linking and social distance}
We first analyze the effect of topological distances on new citation creation by using the plain social distance $d$ and the detachment-based distance $\partial$.
Figure~\ref{prop-soc} depicts the results for the social-distance-based propension $f$; the trend for $f^\partial$ is essentially similar, although not depicted here due to length constraints. On the whole, both propensity profiles are strongly and generally exponentially decreasing with higher distances, reflecting the effect on link creation likelihood of structu\-ral/to\-po\-logical and attentional remoteness: link creation basically occurs in the topological neighborhood, often not much farther than a couple of clicks away. Interestingly, above a certain threshold propensities stop decreasing: in other words, below a certain level of closeness, all bloggers are equally remote.

Specifically in terms of social distance, propensities are about at least one order of magnitude larger at distance $1$ than other distances, for all networks. Links at distance $1$ are actually repeated links, indicating that most relationships, by large, tend to occur between already connected bloggers; then, secondarily, towards friends of friends. Rather than speaking of a ``small-world'', in this case, one would rather talk of a ``narrow-world'' \cite{raux:lien}.
When new links are established outside this close circle, the propensity to cite decreases particularly steeply with respect to social distance. Eventually propensities relative to detachment-based distances, while indicative of weighted attention-related processes, still mostly exhibit the same behavior and confirm these topogical effects.

\begin{figure}[!t]
	\centering
\includegraphics[width=.9\linewidth,height=4cm]
{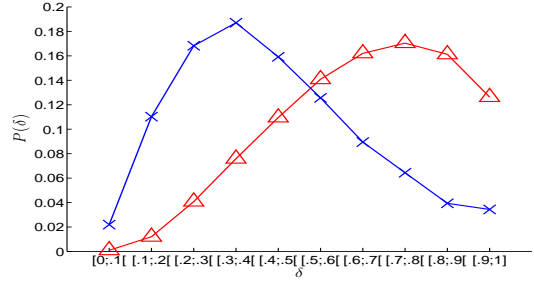}
 	\caption{Semantic distance distributions. {\em Triangles:} distribution computed over the whole set of possible pairs of blogs. {\em Crosses:} distribution computed on pairs of blogs actually linked in the citation network $\bfC$.}
	\label{fig:vois-sem}
 \end{figure}

\subsection{Linking and semantic distance}
Topology thus self-influences topology, yet content distribution may admittedly play a role in further shaping network structure; \cite{adam:poli} demonstrated for instance how partisan divides corresponded to structural ones in the political blogosphere prior to 2004 US elections. 
\begin{figure}[!b]
	\centering
 \includegraphics[width=0.86\linewidth]{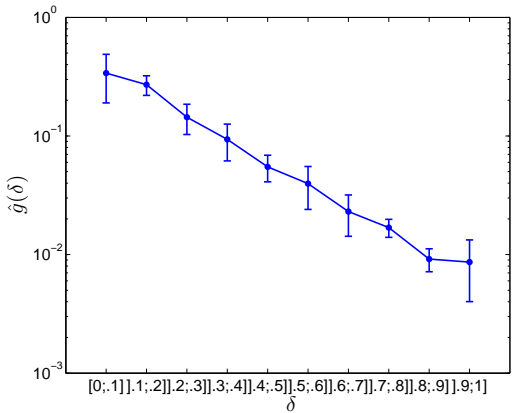}
 	\caption{Propensity for new post citation with respect to semantic distance $\delta$.}
	\label{fig:prop-sem}
 \end{figure}

Here, we can first appraise homophily \emph{statically}, or \emph{a posteriori}, by observing the configuration of links already present  at $t$. We therefore measure the semantic distance $\delta$ between blogs, distinguishing the whole blogosphere from the immediate neighborhood of blogs. We observe on Fig.~\ref{fig:vois-sem} that the immediate neighborhood  is very significantly closer semantically, when compared with the overall semantic distance between pairs of blogs of the whole set $\Bcal$, indicating a very strong \emph{a posteriori} homophily.

This fact suggests a strong homophilic behavior in link creation itself; in other words, it indicates a \emph{dynamic}, or \emph{a priori}, homophily. To check this, 
 we compute propensities for link creation with respect to the semantic distance. The results are plotted on Fig.~\ref{fig:prop-sem} and clearly confirm the above hypothesis. For instance, blogs at a semantic distance less than $.2$ will have a likeliness to cite each other about $10$ times higher than blogs at an average semantic distance and $100$ times than couple of blogs strongly differing semantically.


\subsubsection*{Topological coevolution}

\begin{figure}
	\centering
 \includegraphics[width=0.9\linewidth]{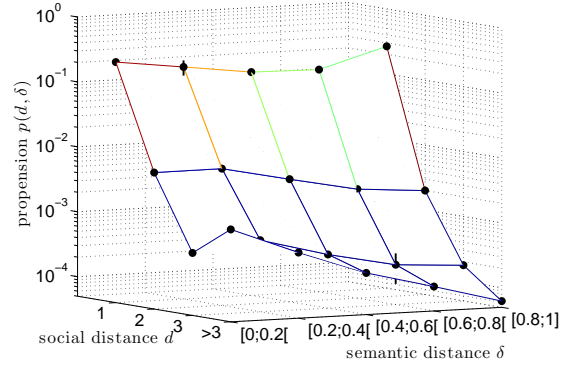}
 	\caption{Two-dimensional propensity with respect to social and semantic distances.}
	\label{fig:prop3d}
 \end{figure}

To appraise how the social and semantic effects mix together, we finally compute propensities in a two-variable setting based on both social and semantic distances,  as shown on  Fig.~\ref{fig:prop3d}. The main conclusion is that, outside of the close circle of repeated citations ($d=1$), the above-mentioned homophilic behavior has a sensible effect, even stronger with increasing social distances. In the case of neighbors however (\hbox{i.e.} repeated citations), the semantic distance has a mixed role. Citations are indeed more likely towards very similar blogs, again ($\delta\in[0;0.2[$), yet, it is also more and even much more likely towards very dissimilar blogs ($\delta\in[0.8;1]$). 


\section{Evolution of Content:\\The Topology-based Dynamics\\of Diffusion}\label{sec:diffusion}\smallskip
Topology thus evolves with respect to content distribution. Yet, in a dual manner, how does the dynamics of content circulation depend on  topological features? 
To assess this, we first need to introduce a notion of diffusion subgraphs (Sec.~\ref{sec:sub}) and some specific characteristics of the underlying citation networks which may be likely to influence the diffusion phenomena, particularly attention-related features (Sec.~\ref{sec:diff}).
  
\subsection{Diffusion subgraph}\label{sec:sub}
More precisely, we focus on explicit diffusion events, which correspond to simultaneously posting some content and referring to another blog which already posted about this same content. We therefore define the notion of {\bf diffusion subgraph}, which gathers every blog which mentioned a given URL in a post, and every directed link $(i,j)$ between these  blogs \emph{such that} $i$ simultaneously mentions the URL and refers to $j$ which had already, previously, mentioned that URL.\\
Technically, given a resource $u\in\Ucal$, we define $\sigma_u$ the \emph{diffusion subgraph of $u$} as a pair of:
\begin{itemize}
\item blogs mentioning $u$ in a post, and
\item  
 directed edges $(i,j)$ of $\bfC$ such that $i$ simultaneously  both cited $j$ and mentioned $u$, \emph{after} $j$ mentioned $u$. \\Formally, these {\bf transmission links} are edges $(i,j)$ such that
$\bfC_t(i,j)>\bfC_{t-1}(i,j)$ (\hbox{i.e.} there is a new link in $\bfC_t$ from $i$ to $j$ at $t$), $\url_t({i,u})=1$ ($i$ mentions $u$ at $t$) and $\exists t'<t$, $\url_{t'}({j,u})=1$ (\hbox{i.e.} $j$ had mentioned $u$ strictly before $t$).
\end{itemize}
We denote such subgraphs $\sigma_u
\,\in\, \mathcal{P}(\Bcal)\times\mathcal{P}(\Bcal\times\Bcal)$.

We say that a diffusion subgraph is \emph{trivial} if its edge set is empty, \hbox{i.e.} if the corresponding URL is not involved in any explicit diffusion event between two blogs.
Of the {$96,637$ URLs of $\Ucal$, only {$11,709$} correspond to non-trivial diffusion subgraphs over the whole collection period. In the remainder, we only focus on these non-trivial subgraphs.

Figure~\ref{fig:diffsubgraph} provides an illustration of a real, non-trivial diffusion subgraph, whose underlying post citation network has previously been illustrated on Fig.~\ref{fig:attention}. In this case, a given URL $u_0$ is first mentioned in blog $a$. 
 It is then mentioned by $c$ on  Feb 19, who cites $a$ on the same day. It then ``diffuses'' to $b$ both from $a$ and $c$ on the next day. Eventually, blog $d$ mentions $u_0$ along with a reference to $b$ on Feb 26.
\begin{figure}
	\centering
\includegraphics[width=0.71\linewidth]{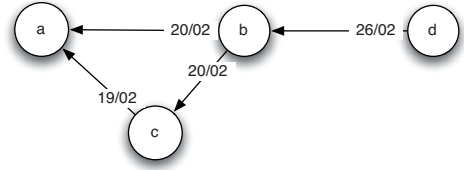}
\caption{Illustration of a diffusion subgraph $\sigma_{u_0}$. Date labels indicate the time when the origin blog both mentioned $u_0$ and did a post citation to the destination.}\label{fig:diffsubgraph}
\end{figure}

\medskip
  
We plotted  on Fig.~\ref{cascadessize} the size distributions of the {$11,709$} non-trivial diffusion subgraphs. Sizes are sensibly heterogeneous both in terms of nodes and links, with a large number of small subgraphs (this observation is consistent with the shape of the cascade size found in  \cite{Leskovec:2007p186}). Most of these subgraphs ($7,016$) consist of a unique transmission event --- $2$ blogs and one link --- while there are {$39,540$} transmission events, over a total of {$229,736$} citation links, \hbox{i.e.} slightly more than one sixth of post citations are also transmission links.

\begin{figure}[!t]
	\centering
\includegraphics[width=.7\linewidth]{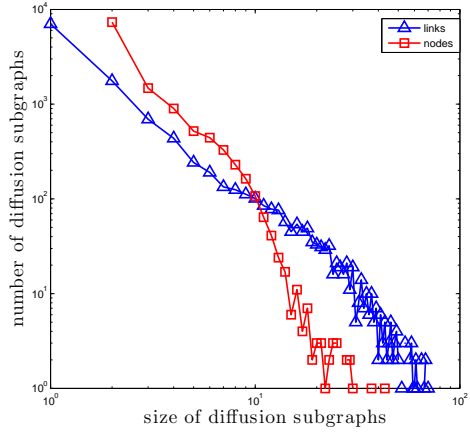}
 	\caption{Size distributions of diffusion subgraphs, in terms of nodes ({\em red squares}) and links ({\em blue triangles}).}
	\label{cascadessize}
 \end{figure}

 \subsection{Diffusion-driven topological features}\label{sec:diff}
\subsubsection{Total attention}
A quite simple ego-centered measure likely to be relevant to study diffusion relates to the notion of {\bf total attention} exerted by a blog $j$, defined as the sum of attentions exerted on all ``attentive'' blogs $i$: $$\bfalpha_t(j) = \sum_i \bfa_t(i,j)$$.

On Fig.~\ref{fig:attention}, the {total attention} $\alpha_t(c)$ exerted by blog $c$  aggregates attentions from blogs $b$, $a$, $d$, $e$ and $f$ towards $c$, it is equal to $2.45$.

\subsubsection{Edge-range distance}
In addition, we now need a notion of structural distance that captures a feeling of remoteness between nodes \emph{already} connected, obviously because the study of explicit diffusion is based on blogs which explicitly link towards and are thereby connected to other blogs. To this end, we use the notion of {\bf edge range}, which has been notably recently used in diffusion studies in \cite{Kossinets:2008p1977} and which had been initially defined in \cite{watt:smal} for a link $(i,j)$ as the distance between $i$ and $j$ if link $(i,j)$ were removed.

\begin{figure}
\centering
\includegraphics[width=.5\linewidth]{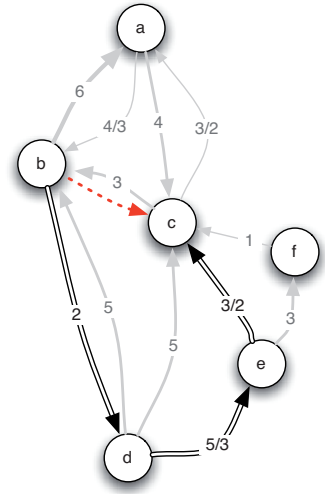}
\caption{\emph{Edge-range calculation:} we compute for instance edge-range $\bfr(b,c)$. 
The link between from $b$ to $c$ is first removed before computing the minimal-cost path from $b$ to $c$, using detachment values $\dk$ computed on $\bfC$. On this example, the path is $(b-d-e-c)$ and we have $\bfr(b,c)=\frac{31}{6}$. Note that paths with less steps such as $(b-a-c)$ may happen to be actually more expansive (with a cost of $10$ in this very case). 
}\label{fig:edge-range}
\end{figure}

We extend this notion to the case of a graph weighted with detachment values. Formally, we define edge range $\bfr(i,j)$ of link $(i,j)$ in the weighted detachment-based graph $\bfD$ as the minimal weighted distance between $i$ and $j$ when link $(i,j)$ is removed.\\
In other terms, it is the minimal sum of detachment values along the ``best'' indirect path from $i$ to $j$; or, so to say, the minimal total attentional cost an information requires to travel from a blog $j$ to $i$ if the edge from $i$ to $j$ were removed. More simply, it is also the detachment-based distance in a graph where edge $(i,j)$ has been removed. \\See an example of edge range calculation on Fig.~\ref{fig:edge-range}.


\subsection{Information relaying and attention} 
The likeliness of a blog to be influent, by inducing content diffusion, is often said to be directly related to the number of links which flows into it \cite{gill:meas,Java:2006p1951} --- influential bloggers being those who have more incoming links or those who have the largest audience. Following this standpoint, one can check the influence of ego by examining how ego-centered measures correlate with actual diffusion.

\begin{figure}
\centering
\includegraphics[width=\linewidth]
{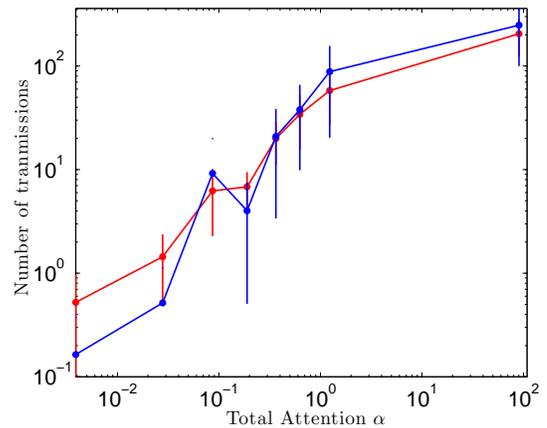}
\caption{Mean number of first (blue dots) and second (red dots) transmission links produced by initiating blogs depending on their total attention $\alpha$.\newline {(NB: distributions are plotted using $8$ quantiles of $\alpha$ values to accommodate for their sensibly heterogeneous spread).}\label{fig:inflcasc}}
\end{figure}

As a first step, we check the correlation between the \emph{total attention} of a blogger using a URL for a first time, and the transmission links s/he induces, \hbox{i.e.} as an \emph{originator} in the corresponding diffusion subgraph.
Figure~\ref{fig:inflcasc} therefore depicts in blue the mean absolute number of such \emph{first} transmission links provoked by blogs having a given total attention $\alpha$.\footnote{Although not depicted here, we found similar correlations between the number of transmission links and the audience size in the broad sense, as measured by the number of \emph{incoming links}. We nonetheless suggest \emph{total attention} measures more precisely audience-related effects as  it considers  individual attentional landscapes, by weighting the number of links the referred blog receives \emph{with the relative importance} it bears for the referring blogger.}

Higher total attention values are indeed correlated with a larger number of transmission events. 
In other words, more ``influential'' blogs seem basically and unsurprisingly to be those with larger active readership, broadly speaking.  However, influence appears to increase more than linearly for \emph{total attention} values in the range of $5\cdot 10^{-2}$ to $1$, compared with total attentions below $5\cdot 10^{-2}$. This suggests that there is an accumulative benefit of having a larger total attention; however, this effect seems to be bounded as it vanishes for even higher values: above a certain threshold, the increase in influence is flatter, although still relatively increasing.
On the whole, this ``broken'' shape suggests that the influence of an initiator, as measured by the number of first transmission links, is not a direct, linear result of attention.



\subsection{Information shortcuts and edge range}
Beyond underlining immediate readership effects, \hbox{i.e.} somehow emphasizing that information transmission through citation is more frequent among regularly cited blogs, this kind of strictly ego-centered indicators is likely to provide little knowledge on a wider picture of information pathways; \hbox{i.e.} of propagation flows in terms of what makes an information propagate \emph{more broadly}, in a \emph{wider} arena.

\subsubsection{Second transmissions}

To explore this, we choose to focus on ``second transmissions'' in diffusion subgraphs. In what precedes, we indeed exhibited that first transmissions were likely to be initiated by blogs having a large \emph{total attention}.
 First transmissions are relative to a given initial source --- i.e. an initiator of a diffusion subgraph, who mentions a resource $u$ without citing another blogger who mentioned $u$ beforehand --- while second transmissions are relative to a blog which already relays a resource. In other terms, it relates to the \emph{longevity} of the diffusion phenomenon. Put simply, once a resource has been transmitted, how likely is it to pursue its way into the blogosphere? 

As can be infered from the red curve on Fig.~\ref{fig:inflcasc}, the effectiveness of second transmissions are determined by the attention of the initiator in roughly the same way as first transmissions were; in other words, attention does not inform us more on the longevity of the informational cascade. 

\subsubsection{Weak ties and edge range}
Rather, information spreading could depend on more holistic features related to the position of the pairs of individuals in the network: consistently with the vast amount of sociological literature on diffusion, information propagation could be more efficient along ``weak ties'' connecting remote areas of a network \citep{roge:new}. In this respect, we use edge range values as they provide a less local information than ego-centered attentional profiles. Higher edge range values are indeed typical of pairs of blogs which would otherwise be relatively far apart within the network, in terms of informational and attentional pathways, if the link between them were absent. 
As such, higher values loosely indicate weak ties \cite{gran:stre}. 


%
 
 
 
In particular, we examine the hypothesis that an information which has been channeled through a weak-tie as a ``shortcut'' may be more ``contagious'' for further diffusion.
{To test this hypothesis, we measure the number of \emph{transmission links} in each diffusion subgraph with respect to the edge range of the edge from which the original resource was cited. In other words, if $i$ is an initiator in subgraph $\sigma_u$, $j$ cites $i$ for $u$, we then examine the number of blogs $k$ in  $\sigma_u$ such that $(k,j)$ are edges of $\sigma_u$, with respect to $\bfr(j,i)$. 
}

 \begin{figure}[t]
	\centering
\includegraphics[width=0.8\linewidth,height=5cm]{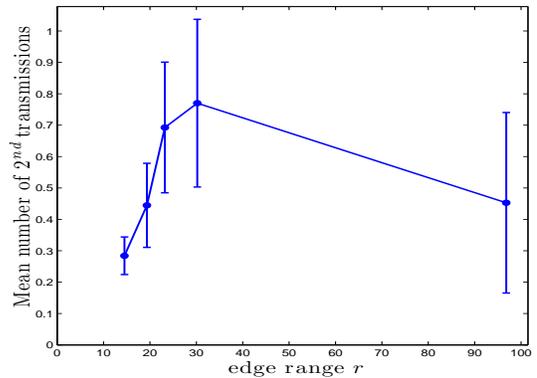}
 	\caption{Mean number of second transmission links $(k,j)$ with respect to the edge range value $\bfr(j,i)$ of the first transmission. Data scarcity led us to bin $\bfr$  into five quintiles.
	\label{fig:weak-tie}}
\end{figure}

{The corresponding statistics, plotted on Fig.~\ref{fig:weak-tie}, shows that resources which transited through edges of higher $\bfr$ generally tend to propagate to a greater number of blogs than for lower $\bfr$. This is however valid below a certain threshold, after which links seem to be too weak to efficiently provoke second transmissions. As such \emph{weak ties}, \hbox{i.e.} with higher edge range, {proportionally act more as catalyzers for ongoing diffusions} in that they connect otherwise relatively remote areas. 

To sum up, blogs (i) connected through a medium edge range to (ii) a ``high attention'' blog realize higher numbers of second transmissions.

}


  
\section{Conclusion}
Social and semantic dimensions are essentially co-determined in this blog network: first, both social and semantic topologies drive new interactions, specifically through a strong homophilic behavior and link creation within the structural neighborhood.
Second, information circulation is shaped both by social and attentional topology, in a broad framework where influence in understood in relatively holistic terms. In particular, we showed how specific structural features may be associated to information pathways: while an ego-centered property such as total attention indicates a higher capacity to disseminate particular online resources, a non-ego-centered property such as edge range indicates that weaker links generally tend to bring richer diffusion in the longer term. Higher attention combined, later on, with higher edge range significantly enhance the capacities for an online resource to be further diffused. 

More broadly, we see this whole framework as a preliminary to a deeper understanding of the joint, coevolving dynamics of social and semantic structures, or the joint evolution of topology and information distribution, notably in the case where both dimensions evolve at comparable timescales.


\small
\section{Acknowledgments}
This work has been partially supported by the French National Agency of Research (ANR) through grant ``\emph{Webfluence}'' \#ANR-08-SYSC-009. We warmly thank Franck Sajous for NLP assistance; 
and RTGI and Guilhem Fouetillou for providing the original dataset and relevant feedback.

\bibliography{../blogs} 
\bibliographystyle{IEEEtran} 
\end{document}